\documentclass[pra,aps,preprint,byrevtex]{revtex4}
\usepackage{epsfig}
\usepackage{amsmath}

\newcommand {\bra}[2] {\mbox{}_{#2}\langle #1 |}
\newcommand {\ket}[2] {| #1 \rangle_{#2}}
\newcommand {\eqn}[1] {Eq.~(\ref{#1})}
\newcommand {\half} {\frac{1}{2}}
\newcommand {\fig}[1] {Fig.~\ref{#1}}
\newcommand {\figwidth} {80mm}

\newcommand {\Sec}[1] {Sec.~\ref{#1}}

\begin{document}
\title{Teleportation using coupled oscillator states}
\author{P.~T. Cochrane}\email{cochrane@physics.uq.edu.au}
\author{G.~J. Milburn}\email{milburn@physics.uq.edu.au}
\author{W.~J. Munro}\email{billm@physics.uq.edu.au}
\affiliation{Centre for Quantum Computer Technology, Department of Physics,
The University of Queensland, QLD 4072, Australia.}

\date{\today}

\begin{abstract}
We analyse the fidelity of teleportation protocols, as a function of
resource entanglement,
for three kinds of two mode oscillator states: states with
fixed total photon number, number states entangled at a beam splitter, and
the two-mode squeezed vacuum state.
We define corresponding teleportation protocols for each case including
phase noise to model degraded entanglement of each resource.
\end{abstract}


\maketitle

\section{Introduction}
\label{sec:intro}

Quantum entanglement plays a central role in the emerging fields of  quantum
computation\cite{DiVincenzo:1995:1,Grover:1997:1,Ekert:1998:1,Jozsa:1998:1,%
Vedral:1998:2},
quantum cryptography\cite{Bennett:1992:1,Kempe:1999:1}, quantum
teleportation\cite{Bennett:1993:1,Bennett:1996:2,Braunstein:2000:2,%
Plenio:1998:3,Braunstein:2000:1,Braunstein:1998:2},
dense coding\cite{Braunstein:2000:3}
and quantum
communication\cite{Bennett:1999:3,Schumacher:1996:1,Schumacher:1996:2}.
The characterisation of entanglement  is a  challenging
problem\cite{Hill:1997:1,Bennett:1996:1,Wootters:1998:1,DiVincenzo:1999:3,%
Vedral:1997:1,Vedral:1998:1} and considerable effort
has been invested in characterising entanglement in a variety of
contexts\cite{Jonathan:1999:1,Jonathan:1999:2,Horodecki:1999:1,Buzek:1997:1,%
Dur:2000:1,Murao:2000:1,Vedral:1999:2}.

One such context, quantum teleportation, has played a crucial role in
understanding how entanglement can be used as a resource
for communication. The recent experimental
demonstrations\cite{Furusawa:1998:1,Boschi:1998:1} suggest that quantum
teleportation
could be viewed as an achievable experimental technique to quantitatively
investigate quantum entanglement. Teleportation is
a way of transmitting an unknown quantum state to a distant receiver with
far better reliability than can be achieved
classically. As the entanglement of the enabling resource is degraded, the
fidelity of the teleportation protocol is
diminished.

In this paper we attempt to make this intuition more precise using specific
examples from quantum optics. Three entangled
resources are considered: states with
fixed total photon number, number states entangled at a beam splitter,
and the two-mode squeezed vacuum state\cite{Milburn:1999:1}.  The
examples we discuss exhibit quantum correlations between the photon number
in each mode and, simultaneously, between the phase
of each mode.

In reality, the entanglement will not be perfect, but degraded to some
extent by uncontrolled interactions with an environment during
formation. To model this we consider phase fluctuations on each mode
independently.  In the
limit of completely random phase we are left with only the classical
intensity (photon number) correlations. The state is no longer entangled
and the fidelity of the
protocol depends only on the classical intensity correlations remaining in
the resource.

\section{Entanglement and Teleportation}
\label{sec:ent}

Intuitively entanglement refers to correlations between distinct subsystems
that cannot be achieved in a classical
statistical model. Of course correlations can exist in classical mechanics,
but entanglement refers to a distinctly
different kind of correlation at the level of quantum probability
amplitudes. The essential difference between quantum and
classical correlations can be described in terms of the separability of
states\cite{Rungta:2000:1,Deuar:2000:2,Pittenger:2000:1,Dur:1999:1,%
Rudolph:2000:1,Horodecki:1997:2,Peres:1996:1,Caves:1999:1}.
A density operator of two subsystems is separable if it can be written as
the convex sum\cite{Peres:1996:1}
\begin{equation}
\rho = \sum_A w_{A} \rho'_{A} \otimes \rho''_{A},
\label{eq:convSum}
\end{equation}
where $\rho_{A}'$ and $\rho_{A}''$ are density matrices for the two 
subsystems and the $w_{A}$ are positive weights satisfying $\sum_A 
w_{A} = 1$.
For example, for two harmonic oscillators ($a$ and $b$) the density operator
which has correlated energy
\begin{equation}
\rho_{ab}=\sum_{n=0}^\infty p_n|n,n\rangle\langle n,n|
\end{equation}
is separable, (where we use the notation
$|n,n\rangle=|n\rangle_a\otimes|n\rangle_b$)
while the pure state \begin{equation}
|\Psi\rangle_{ab}=\sum_{n=0}^\infty\sqrt{p_n}|n,n\rangle
\end{equation}
has the same classical correlation but is not separable. In this form we
see that is possible for a separable and an
entangled state to share similar classical correlations for some variables.

Consider a  communication protocol in which the results of measurements
made on a physical system are transmitted to a distant
receiver. The goal of the receiver is to reconstruct the physical state of
the source, using only local resources, conditioned
on the received information. The communication that takes place is of
course entirely classical. In  a teleportation protocol
there is one additional feature: quantum correlations, entanglement, are
first shared between the sending and receiving
station. The degree of entanglement shared by sender and receiver is called
the teleportation {\em resource}. If there is no
shared quantum correlation between the sender and receiver, the protocol is
called classical.

The extent and nature of the
quantum correlations in the resource determine  the fidelity of the
protocol. Under ideal conditions the unknown state of some
physical system at the transmitting end can be perfectly recreated in
another physical system at the receiving end.  There are many ways
in which actual performance can  differ from the
ideal. In this paper we  analyse the change in the performance of teleportation
protocols as the degree of entanglement in the resource is varied by
decoherence.
Our primary objective is to use the fidelity of a teleportation protocol to
compare and contrast different kinds of entangled oscillator states.

A general teleportation  protocol proceeds as follows. The sender, Alice,
has a {\em
target} state, $\ket{\psi}{T}$,
she wishes to teleport to Bob, the receiver.
Alice and Bob each have access to one part of an entangled bipartite
physical system prepared in the state $\ket{\psi}{AB}$.
In this paper the  bipartite physical system is a two-mode electromagnetic
field. In order to send
the state of the target to Bob, Alice performs a joint measurement on the 
target and her mode.  She then sends the information  
gained from these
measurements to Bob via a classical channel.
Bob performs local unitary transformations on the mode
in his possession according to the information Alice
sends to him, thereby attempting to recreate the initial
target state.  We quantify the quality of the protocol by the probability that
Bob's received state is the same as the target state. This quantity is
known as the
{\em fidelity}.

The fidelity of quantum teleportation protocol is determined by the degree
of shared entanglement,
the quality of the measurements made by the sender, the quality  of the
classical communication channels used  and
 how well Bob can implement the desired unitary transformations.
In this paper we will discuss only the first of
these; the amount of shared entanglement. In the original teleportation
protocol\cite{Bennett:1993:1}, the bipartite system was
made up of two systems each described by a two dimensional Hilbert space,
that is to say, two qubits, and the shared entangled
state was a {\em maximally} entangled state\cite{Bennett:1996:1}. In
the case of two correlated harmonic oscillators,
or two field modes, we cannot define maximal entanglement in quite the same
way, as the entropy of each component system can
be arbitrarily large.  In this paper we define extremal entangled pure
states of two field modes in terms of the total
mean photon number and the total maximum photon number.

In the case of a system with a finite dimensional Hilbert space, a state of
maximum (von Neumann) entropy is simply the identity operator in
that Hilbert space. A natural generalisation of this idea to infinite
Hilbert spaces would define a maximum entropy state
subject to some constraint, such as mean energy or total energy. These of
course define the canonical ensemble and micro
canonical ensemble of statistical mechanics. In the case of entangled pure
states the Araki-Lieb\cite{Wehrl:1978:1} inequality
indicates that the entropy of each component system is equal.
As the entropy of a harmonic oscillator
scales with mean energy, this indicates that each component subsystem has
the same mean energy. If we maximise the entropy
of each subsystem subject to a constraint on the mean energy, the state
must be a thermal state. The entangled pure two mode
state for which the reduced density operator of each mode is thermal, is the
squeezed vacuum state.
\begin{equation}
|\lambda\rangle=(1-\lambda^2)^{1/2}\sum_{n=0}^\infty \lambda^n|n,n\rangle.
\end{equation}
The mean photon number in each mode is given by
$\bar{n}=\lambda^2/(1-\lambda^2)$. If, however, we constrain the total photon
number, $N$, of each mode we get a very different expression for a
maximally entangled state,
\begin{equation}
|N\rangle = \frac{1}{\sqrt{1+N}}\sum_{n=0}^N |N-n,n\rangle.
\end{equation}
The entropy of the reduced state of each mode is $\ln(1+N)$ while the mean
photon number is $N/2$.  While squeezed
vacuum states may be achieved in the laboratory,
states with fixed total photon number have not been produced, and will not
be possible until we have a reliable $N$ photon
source.  There are now a couple of proposals for such
sources\cite{Imamoglu:1994:1,Foden:2000:1,Brunel:1999:1} and it may not
be too long before they are
used in teleportation schemes.

Teleportation fidelity for infinite dimensional Hilbert spaces must
necessarily vary from unity for an arbitrary target state,
as the notion of a maximally entangled resource differs from the finite
dimensional case.
The teleportation protocol can also be degraded
by unknown incoherent processes that corrupt the purity of the shared
entanglement. Of course in some
cases these incoherent processes may destroy the correlations entirely, for
example by absorbing all the photons in each mode
before Alice and Bob get to use them. In this paper, however, we will only
consider those decoherence processes that change the
purity of the states and leave unchanged the classical intensity
correlations.

\section{Ideal Resource}
\label{sec:ideal}

In a recent paper by Milburn and Braunstein\cite{Milburn:1999:1} a
teleportation protocol was presented using joint measurements of the photon
number difference and phase
sum on two field modes.  This protocol is possible because the number
difference and phase sum operators commute, thus  allowing determination of
these quantities simultaneously and to arbitrary accuracy. 

Number \emph{sum} and phase \emph{difference} operators also commute,
implying that if eigenstates of these
operators can be found then a teleportation protocol is possible.
Such a protocol is discussed below.
Recently, teleportation
using number sum and phase difference measurements was
described~\cite{Yu:2000:2}. That work did not  address how
the degree of entanglement in the resource changes the teleportation
fidelity as we do here.

\begin{figure}
\centerline{\epsfig{file=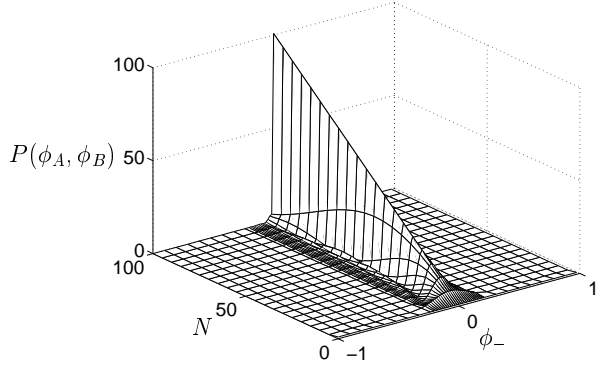, width=\figwidth}}
\caption{Joint phase probability density.  As $N$ increases the
probability density becomes very narrowly peaked about $\phi_{-} =
0$.  The $\phi_{-}$ axis is in units of $\pi$.}
\label{fig:phaseDist}
\end{figure}

Because the number sum
and phase difference operators commute we look for simultaneous eigenstates
of these observables.
Consider states of the form
\begin{equation}
\ket{\psi}{AB} = \sum_{n = 0}^N d_n \ket{N-n}{A}\ket{n}{B}
\label{eq:genIdeal}
\end{equation}
which are eigenstates of number sum with eigenvalue $N$.
The labels $A$ and $B$ refer to the sender's and receiver's component of
the entangled modes
respectively, and the $d_{n}$ satisfy $\sum_{n} |d_{n}|^{2}$.  This state  will 
be maximally  
entangled when the $d_n$ are all equal, giving the resource,
\begin{equation}
\ket{\psi}{AB} = \frac{1}{\sqrt{N+1}} \sum_{n=0}^N \ket{N-n}{A}\ket{n}{B}.
\label{eq:idealRes}
\end{equation}
This state tends towards eigenstates of phase difference as
$N\rightarrow\infty$.
To see this consider the joint phase probability density of
\eqn{eq:idealRes}, which is determined by the ideal joint phase operator
projection operator
$|\phi_A\rangle\langle\phi_A|\otimes|\phi_B\rangle\langle\phi_B|$
as  $P(\phi_{A},
\phi_{B})=tr(\rho_{AB}|\phi_A\rangle\langle\phi_A|\otimes
|\phi_B\rangle\langle\phi_B|)$
where\cite{Walls:1995:1,Susskind:1964:1,Braunstein:1996:1}
\begin{equation}
\ket{\phi}{} = \sum_{n = 0}^\infty e^{in\phi} \ket{n}{}.
\label{eq:phaseState}
\end{equation}
Substituting the state in \eqn{eq:genIdeal} we have
\begin{equation}
P(\phi_{A}, \phi_{B}) = \frac{1}{N+1}\left| \sum_{n=0}^N e^{in\phi_-}\right|^2
\label{eq:phaseDist}
\end{equation}
where $\phi_- = \phi_{A} - \phi_{B}$.  The probability density
as a function of $N$ and $\phi_-$ is shown in \fig{fig:phaseDist}, and
indicates that the density becomes sharply peaked about $\phi_- =
0$ in the interval $[-\pi,\pi]$ as $N$ gets
larger. Hence the states of \eqn{eq:idealRes} tend to eigenstates of
phase difference with increasing $N$.

The state to be teleported, the target state, can be written in the general
form
\begin{equation}
\ket{\psi}{T} = \sum_{m=0}^\infty c_m \ket{m}{T}\ .
\label{eq:target}
\end{equation}
The input state to the protocol is  then
\begin{eqnarray}
\ket{\psi}{} &=& \frac{1}{\sqrt{N+1}}\sum_{m = 0}^\infty \sum_{n =0}^N c_m
\ket{m}{T}\ket{N - n}{A}\ket{n}{B}.
\label{eq:inputIdeal}
\end{eqnarray}
If Alice measures the number sum of the target
and her component of the entangled resource (i.e. $\hat{N}_A+\hat{N}_T$)
with result $q$, the
conditional state of the total system is
\begin{equation}
\ket{\psi^{(q)}}{} = \left [P_I(q)(N+1)\right ]^{-1/2}\sum_{n} c_{q-N+n}
\ket{q-N+n}{T}\ket{N-n}{A}\ket{n}{B},
\label{eq:condQIdeal}
\end{equation}
where $n$ runs from ${\mathrm{max}}(0,N-q)$ to $N$. The probability of
obtaining the result $q$ is
\begin{equation}
P_{I}(q) = \frac{1}{N+1} \sum_{n} |c_{q - N + n}|^2\ .
\label{eq:probIdeal}
\end{equation}
The subscript $I$ emphasises
that this probability refers to the idealised resource.
Alice now measures the phase difference with result $\phi_-$. The
conditional state of Bob's mode is then the pure state
\begin{equation}
\ket{\psi^{(q,\phi_-)}}{B} = \left [P_I(q)(N+1)\right ]^{-1/2}\sum_{n}
e^{2in\phi_-} c_{q-N+n} \ket{n}{B}.
\label{eq:bobIdeal}
\end{equation}
Using the results $q$ and $\phi_-$, and knowledge of the number of
Fock states in the resource ($N$), Bob has sufficient information
to reproduce the target state.  He does this by
amplifying his mode so that $\ket{n}{B} \rightarrow
\ket{q-N+n}{B}$ and phase shifting it by
$e^{-i2n\phi_-}$.  The unitary amplification operation is described in
\cite{Wiseman:1994:1}.
These operations complete the protocol and the
state Bob finally has in his possession is
\begin{equation}
\ket{\psi^{(q)}}{out,B} = \left [P_I(q)(N+1)\right ]^{-1/2}
\sum_{n} c_{q-N+n} \ket{q-N+n}{B}.
\label{eq:outIdeal}
\end{equation}
The
fidelity of this protocol depends on the result $q$ and is
\begin{eqnarray}
F_I(q)  & = & \sum_{n} |c_{q-N+n}|^2\\
  & = & (N+1) P_I(q).
\label{eq:Fideal}
\end{eqnarray}
As the fidelity depends on the result of the number sum measurement it
varies from one run to the next. To obtain
an overall figure of merit for the protocol we define the average fidelity,
\begin{equation}
\bar{F}_I = \sum_{q} F_I(q) P(q)
\label{eq:avFidelity}
\end{equation}
In this case we find
\begin{equation}
\bar{F}_I=(N+1)\sum_{q=0}^\infty P_I(q)^2
\end{equation}
To see how well the teleportation protocol performs, we shall
consider some examples.

Let the target state be a number state,
\begin{equation}
\ket{\psi}{T} = \ket{m}{T},
\label{eq:numberTarget}
\end{equation}
so the only coefficient available is $c_m$, which is one.
We find that the teleportation fidelity is unity, independent of the
measurement of $q$,
because the only term appearing in the summations of both the fidelity and the
probability is that corresponding to $c_m$.
Hence, this protocol works perfectly if the target is a number state.

\begin{figure}
\centerline{\epsfig{file=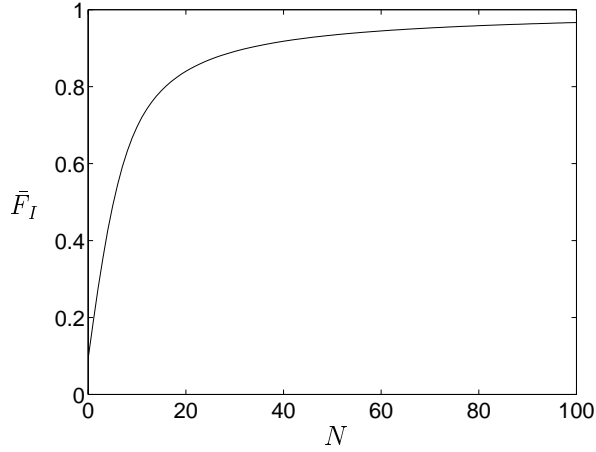, width=\figwidth}}
\caption{Average fidelity as a function of the energy in the ideal resource,
$N$, for a coherent state of amplitude $\alpha=3$.}
\label{fig:Fideal}
\end{figure}

In \fig{fig:Fideal} we show the average fidelity as
a function of the total photon number in the resource for a coherent state
target with
amplitude $\alpha=3$.  It is clear that increasing the number of
photons in the entangled resource improves the
teleportation protocol.

\section{Beam splitter resource}
\label{sec:bsRes}

The resource states discussed in
\Sec{sec:ideal} illustrate the protocol well, but are not
produced by any
known physical interaction.  However, the beam splitter interaction can
be shown to give a resource with similar properties to
\eqn{eq:idealRes}.  The beam splitter interaction is described by
\cite{Sanders:1995:1}
\begin{equation}
\ket{\psi}{AB} = e^{i\frac{\pi}{4}(a^\dagger b + ab^\dagger)}
\ket{N}{A}\ket{N}{B},
\label{eq:BSRes}
\end{equation}
where the operators $a$, $a^\dagger$, $b$ and $b^\dagger$ are the
usual boson annihilation and creation operators for modes $A$ and $B$.
$N$ being the number of photons at each input port of the beam splitter.
Because the number sum of the two modes is a constant ($=2N$) we can rewrite
the resource in terms of eigenstates of number sum.  The resource is
now written as
\begin{equation}
\ket{\psi}{AB} = \sum_{n = 0}^{2N} d_{n-N} \ket{n}{A}\ket{2N-n}{B},
\label{eq:BSResFinal}
\end{equation}
The coefficients $d_{n-N}$ are derived by first using Schwinger's boson
representation of angular
momentum, with total angular momentum quantum number $j=N$, and then
identifying these coefficients as rotation matrix
elements\cite{Sanders:1995:1}.
The $d_{n-N}$ coefficients are
defined by
\begin{equation}
d_{n-N}= e^{-i\frac{\pi}{2}(n-N)} D^N_{n-N,0}(\pi/2)
\label{eq:dndefn}
\end{equation}
where
\begin{eqnarray}
\lefteqn{D_{m',m}^j(\beta) =
\left[(j+m')!(j-m')!(j+m)!(j-m)!\right]^\half}\nonumber\\
& & \times \sum_s \frac{(-1)^{m'-m+s}
\left(\cos\frac{\beta}{2}\right)^{2j+m-m'-2s}
\left(\sin\frac{\beta}{2}\right)^{m'-m+2s}}{(j+m-s)!s!(m'-m+s)!(j-m'-s)!}.
\nonumber\\
&&
\label{eq:dDefnExplicit}
\end{eqnarray}
The variable $s$ ranges over all integer values where the factorials are
non-negative\cite{Biedenharn:1981:1}. It is easy to verify that all
coefficients with $n$ odd are zero.

This protocol proceeds identically to that discussed in
\Sec{sec:ideal}.
We illustrate this variation of the protocol with the pure state form
of the resource as given in \eqn{eq:BSResFinal}.
After a number sum and phase difference 
measurement on modes $T$ and $A$, and then applying the amplification
$\ket{2N-n}{B}
\rightarrow
\ket{q-n}{B}$ and phase shift $e^{-2in\phi_-}$, the output state becomes,
\begin{equation}
\ket{\psi^{(q)}}{out,B} = \left [P_{BS}(q)\right ]^{-1/2} \sum_{n =
0}^{\mathrm{min}(q,2N)}
c_{q-n} d_{n-N} \ket{q-n}{B},
\label{eq:BSOut}
\end{equation}
where the probability for a number sum result $q$ is
\begin{equation}
P_{BS}(q) = \sum_{n = 0}^{\mathrm{min}(q,2N)} |c_{q-n}|^2 |d_{n-N}|^2.
\label{eq:prob_BS}
\end{equation}
The teleportation fidelity is found to be
\begin{equation}
F_{BS}(q) = \frac{1}{P_{BS}(q)} \left|\sum_{n=0}^{\mathrm{min}(q,2N)}
|c_{q-n}|^2 d_{n-N}\right|^2.
\label{eq:FBS}
\end{equation}

\begin{figure}
\centerline{\epsfig{file=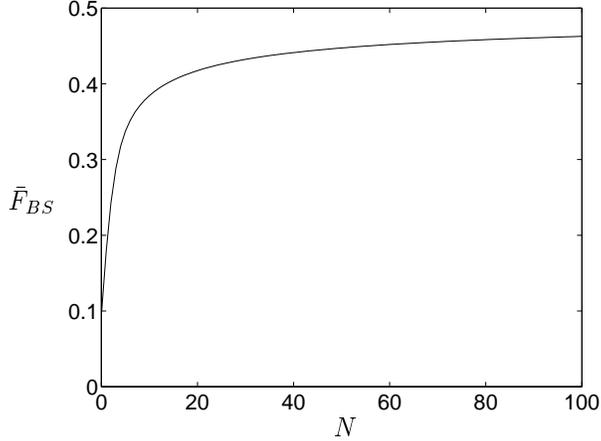, width=\figwidth}}
\caption{Average fidelity as a function of
energy in the beam splitter resource, $N$, for a coherent state of amplitude
$\alpha=3$.}
\label{fig:FBSCohState}
\end{figure}

If we again consider a coherent state target of amplitude
$\alpha=3$, we can compare the beam splitter generated resource with
the ideal resource in \Sec{sec:ideal}.  The average fidelity
as a function of energy in the resource is shown in
\fig{fig:FBSCohState} and is almost identical to
\fig{fig:Fideal} except that its maximum is approximately one half as opposed
to
unity.  This is due to the fact that all terms in \eqn{eq:BSResFinal}
with $n$ odd are zero. Effectively
only half of the perfect correlations in the ideal entangled resource are
available, hence the maximum fidelity we would expect under such circumstances is
$0.5$.
Even so, the state becomes a better teleportation resource  with
increasing $N$ (see \fig{fig:FBSCohState}).

\begin{figure}
\centerline{\epsfig{file=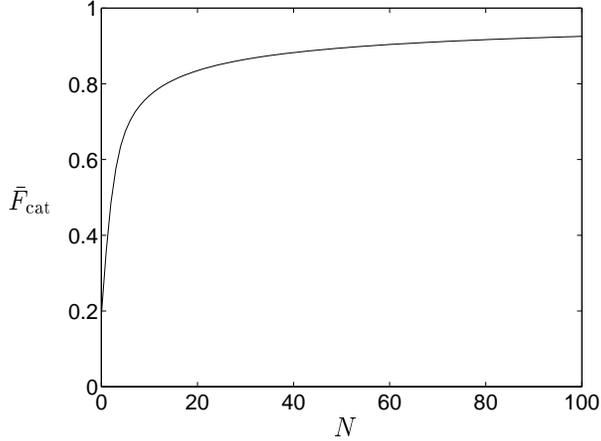, width=\figwidth}}
\caption{Average fidelity as a function of the energy in the beam splitter
resource, $N$, for a ``cat'' state of amplitude $\alpha=3$.}
\label{fig:FBSCatState}
\end{figure}

It is, however, possible to teleport those states
which have no odd photon number components with near unit fidelity.
An example is the  even ``cat'' state, formed from the superposition of
two coherent states of equal  real amplitude but opposite sign
\cite{Cochrane:1999:1},
\begin{equation}
\ket{\psi}{T} = \frac{\ket{\alpha}{T} + \ket{-\alpha}{T}}{\sqrt{2 +
2e^{-2|\alpha|^2}}},
\label{eq:catState}
\end{equation}
  The average fidelity in this case is shown  in \fig{fig:FBSCatState}.
This result implies that it may be possible to tailor resources
for given applications so that certain classes of states may be
teleported well, without necessarily being able to teleport an
arbitrary state.

\section{Decoherence}
\label{sec:noise}

Teleportation requires quantum correlations, in the form of entanglement,
to be shared by the sender and receiver. In this section we
consider how teleportation fidelity changes if decoherence diminishes the
extent of the correlation.
We use a decoherence mechanism, phase diffusion, which does not
change the  intensity (photon number) correlations of the entanglement
resource,
but does destroy the coherence in the number basis.

Phase diffusion is modelled by adding random phase fluctuations to each
mode independently
with the unitary operator
\begin{equation}
U(\theta) \equiv \exp\left[-i(\theta_a a^\dagger a + \theta_b b^\dagger
b)\right],
\end{equation}
where the phase sum and difference $\theta =\theta_a \pm \theta_b$ is
taken to be Gaussian
randomly distributed with a zero mean and variance $\sigma$,
\begin{equation}
P(\theta) = \frac{1}{\sqrt{2\pi\sigma}}
\exp\left(-\frac{\theta^2}{2\sigma}\right).
\label{eq:GaussianForm}
\end{equation}
 Even though a
Gaussian distribution is not periodic it can be taken to be an
approximation of a
true periodic distribution, such as $\cos^{2N}(\theta - \theta_0)$,
which for sufficiently large $N$, is approximately Gaussian near $\theta_0$
with a variance of $1/2N$.

\subsection{Squeezed state resource}
\label{sec:sqNoise}

Reference ~\cite{Milburn:1999:1}
describes a  teleportation protocol using two-mode squeezed vacuum
states as an entanglement resource together with number difference and
phase sum measurements.  The resource for this protocol is
written in the Fock basis as
\begin{equation}
\ket{\psi}{AB} = \sqrt{1 - \lambda^2} \sum_{n=0}^\infty \lambda^n
\ket{n}{A}\ket{n}{B}.
\label{eq:SqRes}
\end{equation}
 The
entanglement between resource modes may be altered by changing the
squeezing parameter, $\lambda$, and by decohering the
resource using phase diffusion.
Applying the phase shift $U(\theta)$
and averaging over all realisations of the phase gives the
resource as a density operator
\begin{eqnarray}
\rho_{AB} &=& (1-\lambda^2) \sum_{n,n'=0}^\infty \lambda^n \lambda^{n'}
e^{-\gamma(n-n')^2} \nonumber\\
&&\times\ket{n}{A}\bra{n'}{}\otimes\ket{n}{B} \bra{n'}{},
\label{eq:SqResNoise}
\end{eqnarray}
where $\gamma=\sigma/2$ describes the degree of decoherence.

The number difference measurement can give a positive or negative  result
and we consider each case separately.
If the state to be teleported is $\rho_T = \sum_{m,m'} c_m c_{m'}^* 
\ket{m}{}\bra{m'}{}$,
the output state at the receiver, conditioned on the positive number
difference, $q$
is
\begin{eqnarray}
\rho_{out,B} &=& \frac{1-\lambda^2}{P_+(q)} \sum_{n,n'=0}^\infty
c_{n+q}
c_{n'+q}^* \nonumber\\
&&\times\lambda^n \lambda^{n'} e^{-\gamma(n-n')^2} \ket{n+q}{B}
\bra{n'+q}{},
\label{eq:SqOutPlus}
\end{eqnarray}
with a corresponding fidelity given by
\begin{equation}
F_{+,\gamma}(q) = \frac{1-\lambda^2}{P_+(q)} \sum_{n,n'=0}^\infty |c_{n+q}|^2
|c_{n'+q}|^2 \lambda^n \lambda^{n'} e^{-\gamma(n-n')^2},
\label{eq:FSqNoisePlus}
\end{equation}
where
\begin{equation}
P_+(q)=(1-\lambda^2)\sum_{n=0}^\infty|c_{n+q}|^2\lambda^{2n}
\end{equation}
is the probability of obtaining a result $q$ for photon number difference
measurements at the sender, which does not depend on the decoherence.

\begin{figure}
\centerline{\epsfig{file=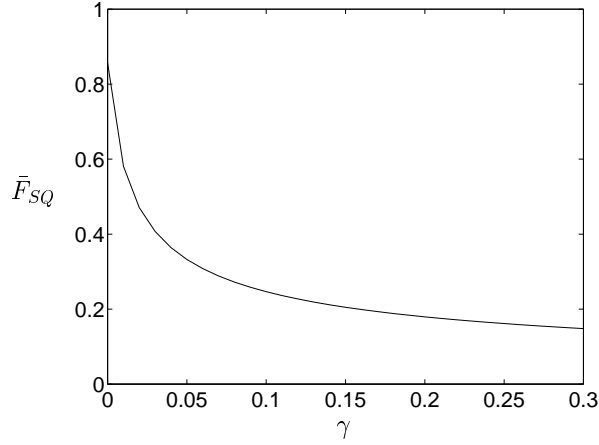, width=\figwidth}}
\caption{Average fidelity as a function of
degree of decoherence, $\gamma$, for a two-mode squeezed vacuum
resource with a squeezing parameter value of
$\lambda=0.8$.  Target is a coherent state of amplitude $\alpha=3$.}
\label{fig:FSq}
\end{figure}

For measurement of negative number difference, $q'=-q$,
The fidelity after teleportation is
\begin{equation}
F_{-,\gamma}(q') = \frac{1-\lambda^2}{P_-(q')} \sum_{m,m'=0}^\infty |c_m|^2
|c_{m'}|^2
\lambda^{m+q'} \lambda^{m'+q'} e^{-\gamma(m-m')^2},
\label{eq:FSqMinus}
\end{equation}
where 
\begin{equation}
P_{-}(q') = (1 - \lambda^{2}) \sum_{m=0}^{\infty} |c_{m}|^{2} 
\lambda^{2(m + q')}.
\label{eq:PSqMinus}
\end{equation}
The average fidelity as a function of degree of decoherence,
$\gamma$, is
shown in \fig{fig:FSq} and behaves as we would expect; decoherence in the
resource
reduces the output quality of the protocol implying that the
entanglement available as a resource for teleportation has decreased.

\subsection{Ideal resource}
\label{sec:idealResNoise}

Applying our decoherence model to \eqn{eq:idealRes} and averaging over
all realisations of the phase we obtain the total state:
\begin{eqnarray}
\rho_{TAB} &=& \frac{1}{N+1} \sum_{m,m' =
0}^\infty \sum_{n,n' = 0}^N
c_m c_{m'}^* e^{-\gamma(n-n')^2} \nonumber\\
&&\times\ket{m}{T}\bra{m'}{}\otimes\ket{N-n}{A}
\bra{N-n'}{}
\otimes\ket{n}{B}
\bra{n'}{}
\label{eq:idealInputNoise}
\end{eqnarray}
where $\gamma$ is the degree of decoherence as before.
After completion of the protocol the fidelity is given by
\begin{equation}
F_{I,\gamma}(q) = \frac{1}{N+1} \frac{1}{P_{I}(q)}
\sum_{n,n'}
|c_{q-N+n}|^2 |c_{q-N+n'}|^2 e^{-\gamma(n-n')^2}.
\label{eq:FidealNoise}
\end{equation}
where $n$ and $n'$ run from ${\mathrm{max}}(0,N-q)$ to $N$ and
$P_{I}(q)$ is given by \eqn{eq:probIdeal}.
It is not difficult to show that by letting $\gamma=0$ we reproduce the result
without noise, \eqn{eq:Fideal}.

\begin{figure}
\centerline{\epsfig{file=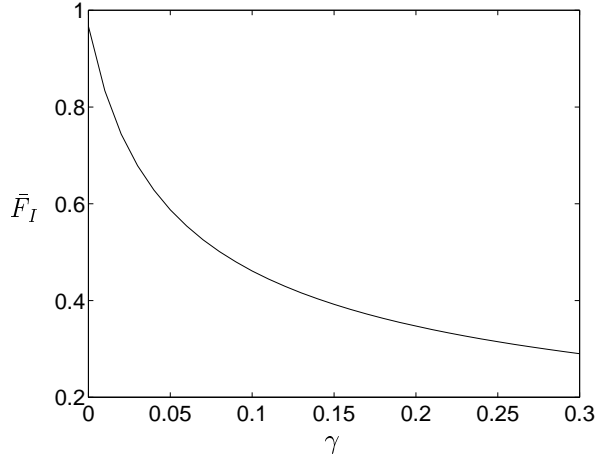, width=\figwidth}}
\caption{Average fidelity as a function of
degree of decoherence, $\gamma$, with an ideal resource energy
corresponding to $N=100$.  Target is a coherent state of amplitude $\alpha=3$.}
\label{fig:FidealNoise2}
\end{figure}

The average fidelity as a
function of the degree of decoherence, $\gamma$, is shown in
\fig{fig:FidealNoise2}
for the example of a coherent state, $\alpha=3$.  As the degree
of decoherence is increased, the fidelity drops away
quickly.  This is because the off-diagonal matrix elements of
$\rho_{AB}$ are being ``washed out''
by the $(n-n')^2$ term in the exponential.  Physically,
we are reducing the entanglement between the
resource modes by making measurement of phase more random and would
expect the ability of the technique
to teleport a state to decrease -- \fig{fig:FidealNoise2} shows this
effect explicitly.

\subsection{Beam splitter resource}
\label{sec:bsResNoise}

We add noise to the beam splitter resource state in the same manner as
described in
\Sec{sec:sqNoise}, obtaining the total state,
\begin{eqnarray}
\rho_{TAB} &=& \sum_{m,m' = 0}^\infty \sum_{n,n' = 0}^{2N} c_m c_{m'}^*
d_{n-N} d_{n'-N}^* e^{-\gamma(n-n')^2}\nonumber\\
& & \times \ket{m}{T}\bra{m'}{}\otimes
\ket{n}{A}\bra{n'}{}\nonumber\\
&&\hspace*{2mm}\otimes
\ket{2N-n}{B}
\bra{2N-n'}{}.
\label{eq:BSInputNoise}
\end{eqnarray}
After the teleportation protocol, we find that the fidelity with
respect to the initial state is given by
\begin{eqnarray}
F_{BS,\gamma}(q) &=& \frac{1}{P_{BS}(q)} \sum_{n,n' = 0}^{\mathrm{min}(q,2N)}
|c_{q-n}|^2 |c_{q-n'}|^2\nonumber\\
&&\times d_{n-N} d_{n'-N}^* e^{-\gamma(n-n')^2}.
\label{eq:FBSNoise}
\end{eqnarray}

\begin{figure}
\centerline{\epsfig{file=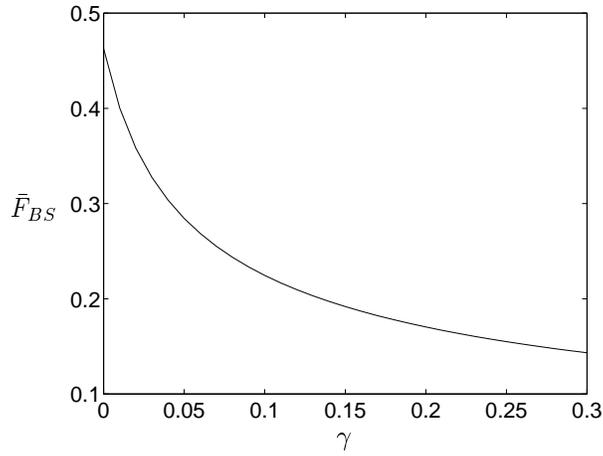, width=\figwidth}}
\caption{Average fidelity as a function of
degree of decoherence, $\gamma$, with a beam splitter resource energy
corresponding to $N=100$.  Target is a coherent state of amplitude $\alpha=3$.}
\label{fig:FBSNoise}
\end{figure}
As we can see in \fig{fig:FBSNoise}, the fidelity decreases due to decoherence
in the resource,
except that the fidelity decreases from approximately one half instead 
of one as in
\Sec{sec:ideal}.

\section{Full decoherence and the classical limit}
\label{sec:infNoise}

Full decoherence corresponds to no entanglement between the resource
modes and a completely flat phase probability distribution.  A flat
phase probability distribution is equivalent to taking
the limit $\gamma\rightarrow\infty$ in the fidelities of
\Sec{sec:noise} thus making the off-diagonal terms in the density
matrix representing the output state, $\rho_{out,B}$, zero.  Physically,
this limit corresponds to
retaining the number correlations but making a measurement of phase
completely arbitrary.  We now suggest that this may be considered a {\em
classical} limit of
the teleportation protocol.

To motivate this point of view, we analyse a classical analogue of the
original qubit teleportation
protocol\cite{Bennett:1993:1}. Consider three classical bits, $T, A,B$,
where $A$ and $B$ are correlated bits shared between
the sender and receiver respectively. The bit labelled $T$ is the target
bit and its state is specified by a
distribution, $p_T(x)$ over the values of the binary variable. The bits
$A$ and $B$ are correlated and have the state
\begin{equation}
p_{AB}(x,y)=\frac{1}{2}\delta_{x,y}
\end{equation}
where $\delta_{x,y}$ is the usual Kronecker delta. The total state of all
three bits is $p_T(z).p_{AB}(x,y)$.
We now  suppose that the sender can measure
the quantity $z\oplus x$ (addition mod 2) on bits $T$ and $A$. The  result
of this measurement is $0$ if both $T$ and $A$ have the same value and $1$
if they have different values. The sender $A$ communicates this result
to the receiver $B$.

The conditional state of the receiver -- given the result of
the measurement, $w$ -- is given  by standard Bayesian conditioning as
\begin{equation}
p_B(y|w)=\frac{\sum_{x,z}p_T(z)p_{AB}(x,y)\delta_{w,z\oplus x}}{P_{TA}(w)}
\end{equation}
where $P_{TA}(w)$ is the probability that the joint measurement on $A$ and
$T$ gives the result $w$. It can be shown that
\begin{eqnarray}
p_B(x|0) &  = & p_T(x)\\
p_B(x|1) & = & p_T(\neg x)
\end{eqnarray}
where $\neg$ is the logical NOT operation.  The receiver $B$ knows the 
result of the joint measurement and can implement  
a local NOT operation if the result of the measurement is $1$.
Given that local operation, we see that the state of the receiver,
$p_B^{out}(x)=p_B(x\oplus w|w)$
is now identical to the state of the target bit, that is to
say it has exactly the same probability distribution.  A little thought
shows the protocol just described is exactly what would be implemented in
the
original qubit protocol if the shared resource between $A$ and $B$ were the
completely decohered
state $\rho_{AB}=(|00\rangle\langle 00|+|11\rangle\langle 11|)/2$. Note
that in this case the only information that can be `teleported' is the
probability distribution for the target bit in the basis in which
$\rho_{AB}$ is diagonal.

For all three entanglement resources considered it can be shown that 
the average fidelity in the fully decohered limit ($\gamma \rightarrow 
\infty$) reduces to:
\begin{equation}
\bar{F}_\infty=\sum_{n=0}^\infty|c_n|^4.
\end{equation}
For example, if the target is a coherent state then this may be shown to 
be
\begin{equation}
\bar{F}(\alpha) = \frac{I_0(2|\alpha|^2)}{e^{2|\alpha|^2}}.
\end{equation}
This is the fidelity between a pure state and a totally mixed state with
the same photon number distribution. We conclude that if the resource
contains only
classical intensity correlations it is only possible to teleport the number
distribution of the target state: no phase information is teleported. In
the sense of
the qubit discussion in the previous paragraph, we call this the classical
limit of the protocol.

\section{Conclusions}
\label{sec:mainConc}

We have shown that a teleportation scheme involving coupled oscillator
states using number sum and phase difference measurements is possible,
given sufficiently large numbers of Fock states in the resource.  The
ability of the scheme to reliably teleport a state was shown to
improve as the number of Fock states in the resource increases.  In
the case of the beam splitter generated resource this physically means
more photons incident on the beam splitter ports.

We have illustrated
the effects of decoherence (in the form of phase diffusion) in three
entanglement resources (ideal, beam splitter generated and squeezed
state) on the fidelity of teleportation and have related this
qualitatively to the change in entanglement of the resource. The
decoherence maintains
the classical intensity correlation inherent in the resource. In the limit
of complete
decoherence the degraded state is only capable of teleporting the number
distribution of the
target state. As this result would have been obtained using standard
Bayesian conditioning
in a classical probabilistic protocol, we argue that it defines a classical
limit for
the quantum scheme.

\begin{acknowledgments}
PTC acknowledges the financial support of the Centre for Laser Science
and the University of Queensland Postgraduate Research Scholarship.
\end{acknowledgments}


\end{document}